\documentclass[11pt,preprint]{aastex}
\usepackage{epsfig}
\newcommand       \beq          {\begin{equation}}
\newcommand       \eeq          {\end{equation}}

\newcommand       \Angstrom     {\,{\rm \AA}} 
          
\newcommand       \cm           {\,{\rm cm}}

\newcommand       \nm           {\,{\rm nm}}
\newcommand       \erg          {\,{\rm erg}}
\newcommand       \eV           {\,{\rm eV}}
\newcommand	  \g		{\,{\rm g}}
\newcommand       \K            {\,{\rm K}}

\newcommand	  \kpc		{\,{\rm kpc}}
\newcommand	  \s		{\,{\rm s}}

\newcommand       \kabs	        {\kappa_{\rm abs}}
\newcommand       \simlt        {\lesssim}
\newcommand       \simgt        {\gtrsim}

\newcommand       \mum          {\,{\rm \mu m}}

\newcommand	  \Teff	        {T_{\rm eff}}
\newcommand       \msun         {\,{\rm M_\odot}}

\newcommand       \simali       {\sim\,}
\newcommand       \rmin         {r_{\rm min}}
\newcommand       \rmax         {r_{\rm max}}
\newcommand       \rstar        {r_\star}

\newcommand       \HD           {{\rm HD\,56126}}

\newcommand       \indxim       {m^{\prime\prime}}
\newcommand       \indxre       {m^{\prime}}
\newcommand       \DT           {\Theta_{\rm D}}
\newcommand  \Eobs {E^{\rm obs}_{\rm 21\mu m}}
\newcommand{\figwidth}{3.6in}
\newcommand{\figwidthL}{4.2in}

\countdef\decade=200
\advance\decade by \year
\countdef\hours=201
\advance\hours by \time
\divide\hours by 60
\countdef\mins=202
\advance\mins by \hours
\multiply\mins by 60
\multiply\hours by 100
\countdef\miltime=203
\advance\miltime by \hours
\advance\miltime by \time
\advance\miltime by -\mins

\shorttitle{On FeO Nanoparticles in Post-AGB Stars}

\begin{document}

\title{
 \vspace*{-2.0em}
  {\normalsize\rm Accepted for publication in 
                  {\it The Astrophysical Journal}}\\
 \vspace*{1.0em}
On Iron Monoxide Nanoparticles as a Carrier of 
the Mysterious 21$\mum$ Emission Feature 
in Post-Asymptotic Giant Branch Stars
	 }
\author{Aigen Li\altaffilmark{1},
        J.M.~Liu\altaffilmark{2}, and
        B.W.~Jiang\altaffilmark{2,1}}
\altaffiltext{1}{Department of Physics and Astronomy,
                 University of Missouri,
                 Columbia, MO 65211, USA;
                 {\sf lia@missouri.edu}
                 }
\altaffiltext{2}{Department of Astronomy,
                 Beijing Normal University,
                 Beijing 100875, China;
                 {\sf bjiang@bnu.edu.cn}
                 }

\begin{abstract}
A prominent, mysterious emission feature peaking at 
$\simali$20.1$\mum$ --- historically known 
as the ``21$\mum$'' feature ---
is seen in over two dozen Galactic and Magellanic Cloud 
carbon-rich post-asymptotic giant branch (post-AGB) stars.
The nature of its carrier remains unknown 
since the first detection of the 21$\mum$ feature in 1989. 
Over a dozen materials have been 
suggested as possible carrier candidates. 
However, none of them has been accepted:
they either require too much material
(compared to what is available in the circumstellar
shells around these post-AGB stars), 
or exhibit additional emission features which are not 
seen in these 21$\mum$ sources.
Recently, iron monoxide (FeO) nanoparticles seem to be a promising 
carrier candidate as Fe is an abundant element and 
FeO emits exclusively at $\simali$21$\mum$. 
In this work, using the proto-typical protoplanetary nebula
HD\,56126 as a test case, we examine FeO nanoparticles 
as a carrier for the 21$\mum$ feature 
by modeling their infrared emission,
with FeO being stochastically heated
by single stellar photons.
We find that FeO emits too broad a 21$\mum$ feature 
to explain the observed one and the Fe abundance required
to be locked up in FeO exceeds what is available in HD\,56126.
We therefore conclude that FeO nanoparticles are unlikely
responsible for the 21$\mum$ feature.
\end{abstract}

\section{Introduction\label{sec:intro}}
During the late stages of evolution,
low- and intermediate-mass 
($0.8\msun$\,$<$\,$M$\,$<$\,8$\msun$) stars undergo
a rapid transition phase of several thousand years 
between the asymptotic giant branch (AGB) phase 
and the planetary nebula (PN) phase. 
Objects in this short-lived stage of evolution
are known as protoplanetary nebulae (PPNe),
a term which is often used interchangeably with 
``post-AGB stars''.

Dust is a general phenomenon of PPNe, as revealed
by its thermal infrared (IR) emission continuum and
spectral features. In recent years, much attention has 
been paid to the so-called ``21$\mum$ feature''. 
This prominent, broad, mysterious emission feature, 
with a peak wavelength at $\simali$20.1$\mum$ 
and a FWHM of $\simali$2.2--2.3$\mum$, 
was first detected in four PPNe (Kwok et al.\ 1989).
To date, it has been seen in 18 Galactic objects 
(Cerrigone et al.\ 2011) and 9 Magellanic Cloud objects 
(Volk et al.\ 2011), all of which are exclusively PPNe.
The 21$\mum$ feature exhibits little shape variation 
among different sources. 
The 21$\mum$ sources exhibit quite uniform
characteristics: they are metal-poor, 
carbon-rich F and G supergiants 
with strong IR excess and over abundant
s-process elements (see Jiang et al.\ 2010).

The carrier of the 21$\mum$ feature remains unidentified, 
although over a dozen candidate materials have been proposed.
Zhang et al.\ (2009a) examined nine inorganic carrier candidates 
for the 21$\mum$ feature, including nano TiC, fullerenes with Ti, 
SiS$_{2}$, doped SiC, silicon and carbon mixture,
SiO$_2$-coated SiC, and iron oxides 
(FeO, Fe$_{2}$O$_{3}$ and Fe$_{3}$O$_{4}$). 
They found that except FeO nanoparticles, 
they are all problematic:
they either require too much dust material 
(compared to what would be available in 
the 21$\mum$ sources)
or produce extra features which are not seen 
in the spectra of the 21$\mum$ sources.

As originally proposed by Posch et al.\ (2004),
FeO (iron monoxide or w\"ustite) nanoparticles
seem to be a promising candidate carrier for 
the 21$\mum$ feature for three reasons:
(1) Fe is an abundant element; 
(2) FeO has a pronounced spectral feature 
    around 21$\mum$; and
(3) except the 21$\mum$ feature, FeO does not 
    have any other notable spectral features.
Posch et al.\ (2004) further argued that FeO nanoparticles
could form and survive in the C-rich shells around
the 21$\mum$ sources (see \S4.4.2 and Footnotes 7,8
of Zhang et al.\ 2009a; also see Begemann et al.\ 1995), 
provided that they are composed of $<$\,10$^3$ atoms 
and have a size of $a$\,$\simlt$\,1\,nm.
Iron oxides (particularly maghemite $\gamma$-Fe$_2$O$_3$ 
and magnetite Fe$_3$O$_4$) have also been suggested as
a potential dust component in the interstellar medium 
(Jones 1990, Draine \& Hensley 2013). 
 
Posch et al.\ (2004) fitted the observed 21$\mum$ emission
feature with FeO of steady-state temperatures in thermal
equilibrium with the stellar radiation field. However, 
with fewer than $\simali$1000 atoms, FeO nanoparticles 
will be stochastically heated by single stellar photons
as their heat capacities are smaller than or comparable 
to the energy of the stellar photons that heat them
(see \S\ref{sec:discussion}). 
Therefore, they will not attain an equilibrium temperature; 
instead, they will experience transient ``temperature spikes''
and undergo ``temperature fluctuations''
(see Draine \& Li 2001).
The stochastic heating of FeO nanoparticles by
individual stellar photons will result in a distribution
of temperatures and consequently, the emission spectra  
are expected to be broader than that from 
a single equilibrium temperature.
 
Zhang et al.\ (2009a) recognized the stochastic heating 
nature of FeO nanoparticles in PPNe. But they did not 
model the IR emission spectra of FeO nanoparticles;
instead, they focused on the abundance constraint on Fe:
they estimated the amount of Fe required to be locked up
in FeO to account for the total emitted power of 
the 21$\mum$ feature by assuming that FeO nanoparticles 
emit at the peak temperature to which they are heated 
upon absorption of a typical stellar photon. 

With an aim at examining the hypothesis of FeO nanoparticles 
as a carrier of the 21$\mum$ feature, we model the vibrational
excitation and radiative relaxation of FeO nanoparticles 
in a proto-typical PPN 
-- HD\,56126 (see \S\ref{sec:results}) -- 
and then compare their model emission spectra 
with the observed 21$\mum$ feature.
In \S\ref{sec:opct} we discuss the optical properties of FeO.
Their heat capacities are discussed in \S\ref{sec:heatcap}. 
In \S\ref{sec:results} we carry out calculations 
for the temperature probability distribution functions
and the emergent IR emission spectra of FeO nanoparticles.
\S\ref{sec:discussion} discusses the results and
summarizes the major conclusions.

\section{Optical Properties of FeO\label{sec:opct}}
To model the heating and cooling of FeO in PPNe,
we need to calculate the absorption cross sections 
of FeO from the ultraviolet (UV) to the far-IR.
This requires the knowledge of the optical properties 
of FeO. The optical properties of FeO vary with 
temperature $T$ as the density of free charge carriers 
decreases with $T$.
Since FeO nanoparticles will have a distribution of
temperatures, we will consider their refractive indices 
at a range of temperatures. 

Henning et al.\ (1995) measured the refractive indices 
of FeO at $T$\,=\,300\,K in the wavelength range of 
$\lambda=0.2\mum$ to $\lambda=500\mum$. 
Henning \& Mutschke (1997) extended the same measurements
to $T$\,=\,10, 100, and 200\,K, but for a smaller wavelength
range of $\lambda=2\mum$ to $\lambda=500\mum$.


\begin{figure}[ht]
\begin{center}
\epsfig{
        file=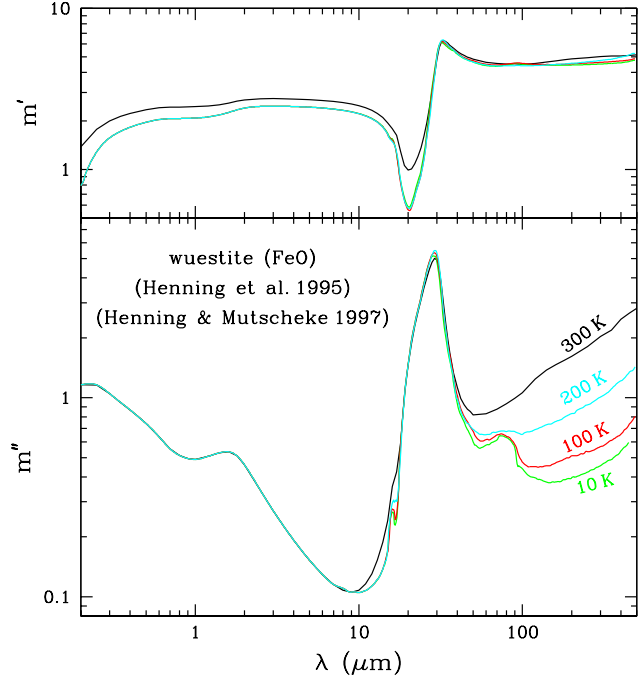,
        width=\figwidth,angle=0}
\end{center}\vspace*{-1em}
\caption{\footnotesize
        \label{fig:nk}
        Optical constants $\indxre$ (upper panel) 
        and $\indxim$ (lower panel) 
        of FeO at $T$\,=\,10\,K (green), 
        100\,K (red), 200\,K (cyan), and 300\,K (black).
        }
\end{figure}

For $T$\,=\,300\,K, we adopt the refractive indices of FeO of 
Henning et al.\ (1995) from $\lambda=0.2\mum$ to $\lambda=500\mum$
(see Figure~\ref{fig:nk}).
For $T$\,=\,10, 100, and 200\,K, the following synthetic 
approach is taken: for $\lambda=2\mum$ to 500$\mum$,
we adopt the imaginary parts ($\indxim$) of 
the FeO refractive indices of Henning \& Mutschke (1997)  
measured at $T$\,=\,10, 100, and 200\,K;\footnote{%
   The $\indxim$ data of FeO of Henning \& Mutschke (1997) 
   for $T$\,=\,10, 100, and 200\,K are rather noisy 
   at $10\mum < \lambda < 15\mum$ and at $\lambda>70\mum$.
   We have smoothed these data before we apply
   the Kramers-Kronig relation. 
   }
for $\lambda<2\mum$ 
we adopt the imaginary parts $\indxim$ of FeO 
of Henning et al.\ (1995) measured at $T$\,=\,300\,K, 
and then smoothly connect the $\indxim$ data at $\lambda<2\mum$ 
to that at $\lambda$\,=\,2--500$\mum$.\footnote{%
   The optical properties of FeO at $\lambda<2\mum$
   are not sensitive to $T$ as the free charge carriers
   (of which the densities are sensitive to $T$)
   mainly contribute to the far-IR continuum.
   }
The Kramers-Kronig relation is then used to
derive the real parts ($\indxre$)
of the indices of refraction of FeO at 
$T$\,=\,10, 100, and 200\,K
from $\lambda=0.2\mum$ to $\lambda=500\mum$.
Figure~\ref{fig:nk} shows the resulting
refractive indices ($m$\,=\,$\indxre$\,+\,$i$\,$\indxim$)
of FeO at $T$\,=\,10, 100, 200, and 300\,K. 

We note that, ideally, we should obtain 
the optical constants of FeO over a much wider
wavelength range, from X-rays to millimeter wavelengths.
However, the lack of experimental $\indxim$ data at
$\lambda<0.2\mum$ and $\lambda>500\mum$ prevents us
from achieving a complete set of ($\indxre$, $\indxim$) data.
Fortunately, for the present study this is not important:
the stellar radiation of PPNe mostly peaks at the visible
wavelength range, and FeO nanoparticles mostly emit their 
energy at $\lambda \sim 20\mum$; therefore, 
the $m$\,=\,$\indxre$\,+\,$i$\,$\indxim$ data 
spanning the wavelength range of 
$0.2\mum < \lambda <500\mum$ are sufficient
for the present study.

\section{Thermal Properties of FeO\label{sec:heatcap}}
FeO nanoparticles consist of several hundred atoms:
with a mass density of $\rho\approx5.7\g\cm^3$,
a FeO grain of spherical radius of $a$ has 
$N_{\rm atom}\approx 400\left(a/{\rm nm}\right)^3$ atoms.
The vibrational degrees of freedom of FeO nano grains,
$3\left(N_{\rm atom}-2\right)$,
are so small that a single stellar photon of energy $h\nu$   
is capable of appreciably raising their temperatures from 
$T_i$ to $T_f$: $\int_{T_i}^{T_f} C(T)\,dT = h\nu$,
where $h$ is the Planck constant, 
$\nu$ is the photon frequency,
and $C(T)\propto N_{\rm atom}$ is
the specific heat of FeO.
At low temperatures 
(i.e., $T\ll \DT$, where $\DT$ is the Debye temperature),
the specific heat is proportional to $T^3$:
$C(T)\,=\,\left(12\pi^4/5\right) N_{\rm atom}\,k\left(T/\DT\right)^3$,
where $k$ is the Boltzmann constant.
As a prior, it is not clear if the condition 
of $T\ll\DT$ will always be met for nano FeO in PPNe.
Therefore, we shall not adopt this simple formula
for $C(T)$; instead, we will adopt a three-dimensional
Debye model with $\DT\approx 430\K$ which closely
reproduces the experimental specific heat of FeO 
measured by Gr{\o}nvold et al.\ (1993) and
St{\o}len et al.\ (1996):
\begin{equation}
C(T) = 3(N_{\rm atom}-2)k f_{3}^\prime(T/\DT)~~,
\end{equation}
\begin{equation}
f_{3}(x) \equiv \int_0^1 \frac{3 y^3 dy}{\exp(y/x)-1} ~~~,~~~
f_{3}^\prime(x)\equiv \frac{d}{dx}f_{3}(x) ~~~.
\end{equation}

\begin{figure}[ht]
\begin{center}
\epsfig{
        file=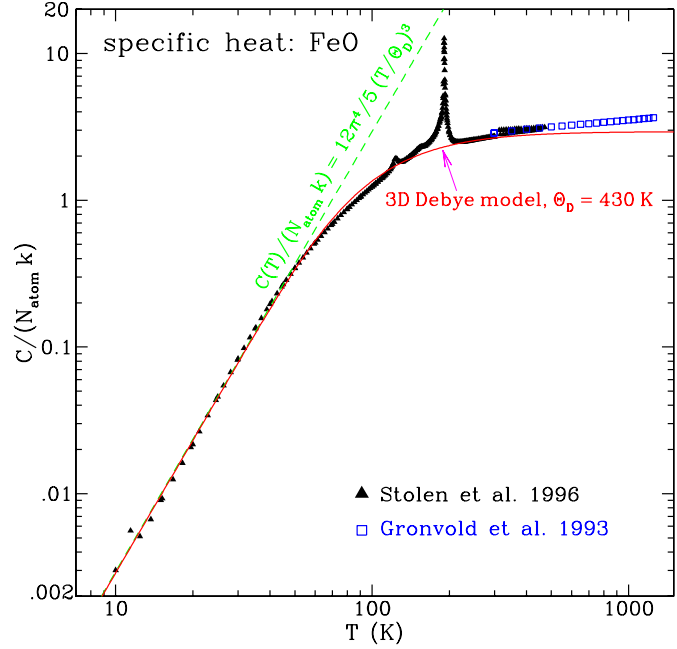, 
        width=\figwidth,angle=0}
\end{center}\vspace*{-1em}
\caption{\footnotesize
        \label{fig:heatcap}
        Comparison of the experimental specific heat 
        of FeO measured by 
        Gr{\o}nvold et al.\ (1993; blue open squares) 
        and St{\o}len et al.\ (1996; black filled triangles)
        and the specific heat given by
        the 3-dimensional Debye model 
        with $\DT=430\K$ (red solid line)
        and the low-temperature approximation
        of $C(T)\propto \left(T/430\K\right)^3$
        (green dashed line).
        The peak at $T\approx 191\K$ is due to
        the magnetic order-disorder transition.
        }
\end{figure}

In Figure~\ref{fig:heatcap} we show the experimental 
specific heat of FeO measured by 
Gr{\o}nvold et al.\ (1993) for $298\K <T <1250\K$
and by St{\o}len et al.\ (1996)
for $10\K <T <450\K$.
Also shown are the 3-dimensional Debye model fit
(with $\DT=430\K$) and the low-temperature approximation
of $C(T)\propto \left(T/\DT\right)^3$.

\section{Results\label{sec:results}}
Let $\kabs(\lambda)$\,=\,$C_{\rm abs}(a,\lambda)/m(a)$
be the mass absorption coefficient of FeO 
at wavelength $\lambda$, where $C_{\rm abs}(a,\lambda)$
is the absorption cross section of FeO of size $a$
at wavelength $\lambda$, and $m(a)$ is the mass of
FeO of size $a$. For FeO nanoparticles in the IR,
$\kabs(\lambda)$ is independent of $a$ since they are
in the Rayleigh regime (i.e., $a\ll\lambda$).
The IR emissivity per unit mass
(in unit of $\erg\s^{-1}\cm^{-1}\g^{-1}$)
from FeO nanoparticles of size $a$ located at
a distance of $r$ from the illuminating star is
\begin{equation}\label{eq:j_lambda}
j_{{\lambda}}(r,a) = \kabs(\lambda)
\int^{\infty}_{0} dT\ 4\pi B_{\lambda}(T)\ dP(r,a,T)/dT ~~,
\end{equation}
where $B_{\lambda}(T)$ is the Planck function 
of temperature $T$ at wavelength $\lambda$,
$dP(r,a,T)$ is the probability that the temperature 
of a FeO nanoparticle of size $a$ at a distance of
$r$ from the illuminating star will be in $[T,T+dT]$. 
The total IR emissivity per unit mass
from FeO nanoparticles of size $a$ is obtained 
by integrating over the entire dust shell
\begin{equation}\label{eq:j_lambda}
j^{\rm tot}_{{\lambda}}(a) = 
\int^{\rmax}_{\rmin}dr\ 
j_{{\lambda}}(r,a)\ 4\pi r^2\ dn(r)/dr ~~,
\end{equation}
$\rmin$ and $\rmax$ are respectively
the inner and outer edge of the dust shell,
and $dn(r)/dr$ is the FeO dust spatial distribution.
The power output per unit mass in the 21$\mum$ band is
calculated by integrating $\Delta j^{\rm tot}_\lambda(a)$
over the entire band, where $\Delta j^{\rm tot}_\lambda(a)$
is the continuum-subtracted $j^{\rm tot}_\lambda(a)$
\begin{equation}\label{eq:Etot}
P_{21\mu{\rm m}}(a) = \int_{21\mu{\rm m}\,{\rm band}}
         \Delta j^{\rm tot}_\lambda(a)\,d\lambda ~~.
\end{equation}
The FeO mass $M({\rm FeO})$ required to 
account for the observed 21$\mum$ emission is
\begin{equation}\label{eq:mfeo}
M({\rm FeO}) =
E_{21\mu{\rm m}}^{\rm obs}/P_{21\mu{\rm m}}(a) ~~,
\end{equation}
where $E_{21\mu{\rm m}}^{\rm obs}$
(in unit of $\erg\s^{-1}$) is the total power 
emitted from the 21$\mum$ feature.
Let $M_{\rm H}$ be the total H mass in the shell.
The Fe abundance (relative to H) required to be
locked up in FeO is
\begin{equation}\label{eq:Fe2H}
\left[{\rm Fe/H}\right]_{\rm FeO} =
M({\rm FeO})/\left[\mu_{\rm FeO}\,M_{\rm H}\right]~~,
\end{equation}
where $\mu_{\rm FeO}=72$ is the molecular weight of FeO.

We take HD\,56126, a proto-typical 21$\mum$ source,
as a test case. 
$\HD$ ($\equiv$\,IRAS07134+1005),
a bright post-AGB star with a spectral type of F0-5I
and a visual magnitude of $\simali$8.3, 
is one of the four 21$\mum$ sources originally discovered 
by Kwok et al.\ (1989). Mid-IR imaging of this object 
at 11.9$\mum$ shows that its circumstellar dust is confined
to an area of 1.2$^{\prime\prime}$--2.6$^{\prime\prime}$ from
the star (Hony et al.\ 2003). Detailed modeling of its dust IR
spectral energy distribution suggested a $dn(r)/dr \sim 1/r$
dust spatial distribution at 
$1.2^{\prime\prime} < r < 2.6^{\prime\prime}$.
Following Hony et al.\ (2003), 
we will adopt a distance of $d\approx 2.4\kpc$ to the star
(and therefore $\rmin\approx 4.3\times 10^{16}\cm$,
$\rmax\approx 9.3\times 10^{16}\cm$),
a stellar radius of $\rstar\approx 49.2\,r_\odot$
($r_\odot$ is the solar radius),
a stellar luminosity of $L_\star\approx 6054\,L_\odot$
($L_\odot$ is the solar luminosity),
and approximate the $\HD$ stellar radiation 
by the Kurucz (1979) model atmospheric spectrum 
with $\Teff = 7250\K$ and $\log g=1.0$.
These parameters determine the starlight intensity
which vibrationally excites FeO nanoparticles.

We adopt the ``thermal-discrete'' method developed
by Draine \& Li (2001) to treat the stochastic heating
of FeO nanoparticles. 
The necessity to model the stochastic excitation
of FeO nanoparticles in the dust shell around HD\,56126
will be justified in \S\ref{sec:discussion}.
For FeO dust of given size $a$
at a given distance of $r$ from the star, 
$P(r,a,T)$ -- the temperature probability distribution 
function -- is calculated from the ``thermal-discrete'' method. 
We first consider spherical dust and use
Mie theory to calculate the absorption cross section
$C_{\rm abs}(a,\lambda)$ of FeO of radius $a$
at wavelength $\lambda$. We adopt the refractive indices
of FeO of $T$\,=\,100\,K (see Figure~\ref{fig:nk}).
As will be seen below, this is justified because 
for nano FeO the temperature probability distribution
peaks around $\simali$100\,K.  

\begin{figure}[ht]
\begin{center}
\epsfig{
        file=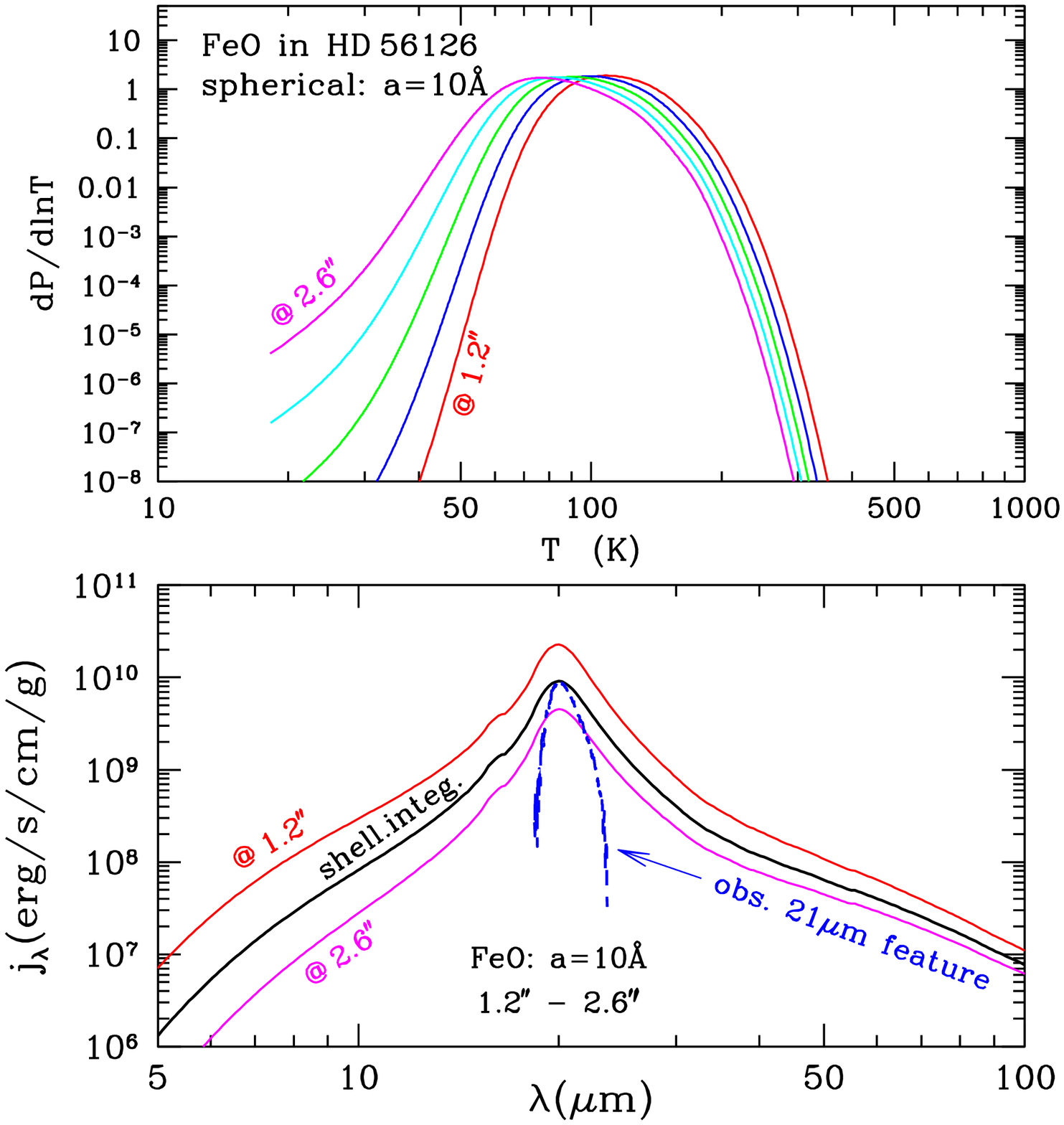,
        width=\figwidth,angle=0}
\end{center}\vspace*{-1em}
\caption{\footnotesize
        \label{fig:irem10A}
        Upper panel: the temperature probability
        distribution functions of spherical FeO 
        of radius $a$\,=\,1\,nm in HD\,56126 
        at various distances 
        (red: 1.20$^{\prime\prime}$,
         blue: 1.46$^{\prime\prime}$,
         green: 1.77$^{\prime\prime}$,
         cyan: 2.14$^{\prime\prime}$, and
         magenta: 2.60$^{\prime\prime}$) 
         from the star.
         Lower panel: the IR emissivity per unit mass
         of FeO of radius $a$\,=\,1\,nm at 
         1.20$^{\prime\prime}$ (red)
         and 2.60$^{\prime\prime}$ (magenta) from the star,
         and that integrated over the entire shell
         (thick black). 
         Also shown is the observed 21$\mum$ emission feature 
         of HD\,56126 (blue dashed; Volk et al.\ 1999) 
         which is scaled to the shell-integrated model spectrum.
         }
\end{figure}

Figure~\ref{fig:irem10A} shows 
the temperature probability distribution functions
$P(r,a,T)$ for spherical FeO of radius $a$\,=\,1\,nm at
a distance of 
1.20$^{\prime\prime}$,
1.46$^{\prime\prime}$,
1.77$^{\prime\prime}$,
2.14$^{\prime\prime}$, and
2.60$^{\prime\prime}$ 
from the star.
The $P(r,a,T)$ distribution functions are broad,
confirming that nano FeO undergoes temperature
excursions in PPNe [otherwise $P(r,a,T)$ should
be strongly peaked and approximated by a delta function].
At a larger distance from the star, $P(r,a,T)$
becomes broader because of the reduced starlight
intensity which leads to a smaller photon absorption rate
and therefore enhances the single-photon heating effect. 
We also note that for $a\simgt1\nm$
$P(r,a,T)$ roughly peaks at $T\sim100\K$
(although the distribution function $P$ 
is still appreciably broad for $a<3\nm$),
implying that the absorbed photon energy of FeO 
will be mostly radiated away
at $T\sim100\K$. This justifies the choice of
the ($\indxre$, $\indxim$) data of FeO of $T\sim100\K$.  
%

\begin{figure}[ht]
\begin{center}
\epsfig{
        file=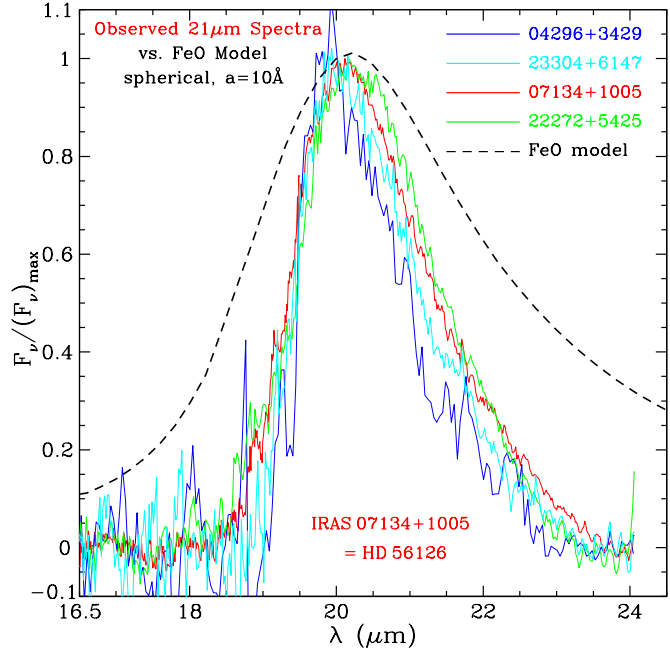,
        width=\figwidth,angle=0}
\end{center}\vspace*{-1em}
\caption{\footnotesize
        \label{fig:Obs}
         Comparison of the shell-integrated 
         21$\mum$ emission feature 
         calculated from spherical FeO of $a$\,=\,10$\Angstrom$
         in the dust shell around HD\,56126 
         (black dashed line; see the thick black line
          in Figure~\ref{fig:irem10A})
         with that observed in four C-rich PPNe 
         (Volk et al.\ 1999): 
         IRAS\,04296\,+\,3429 (blue), 
         IRAS\,22272\,+\,5425 (green),
         IRAS\,23304\,+\,6147 (cyan),
         and IRAS\,07134\,+\,1005 ($\equiv$\,HD\,56126; red).
         The 21$\mum$ feature predicted from FeO is too
         broad to explain the observed feature.
         }    
\end{figure}

Figure~\ref{fig:irem10A} also shows the IR emissivity
per unit mass $j_\lambda(r,a)$ of spherical FeO of radius
$a$\,=\,1\,nm at 1.20$^{\prime\prime}$ and 2.60$^{\prime\prime}$,
as well as that integrated over the entire shell 
$j^{\rm tot}_\lambda(a)$.\footnote{%
   Following Hony et al.\ (2003), we take
   $dn(r)/dr\propto 1/r$.
   }
%
It is apparent that the model feature, with a FWHM of 
$\gamma_{\rm 21\mu m}\approx 3.7\mum$,
is too broad to explain the 21$\mum$ feature of HD\,56126
which has a FWHM of $\simali$2.2$\mum$
(see Footnote-1 of Zhang et al.\ 2009a).
For illustration, we compare in Figure~\ref{fig:Obs}
the model spectrum of FeO of $a$\,=\,10$\Angstrom$ 
with the 21$\mum$ emission spectra observed in
four C-rich PPNe (Volk et al.\ 1999): 
IRAS\,04296\,+\,3429, 
IRAS\,22272\,+\,5425,
IRAS\,23304\,+\,6147,
and IRAS\,07134\,+\,1005 (i.e., HD\,56126),
with each spectrum normalized to its peak value.
Figure~\ref{fig:Obs}, in linear abscissa and ordinate, 
clearly shows that the FeO model results in too broad
a 21$\mum$ feature to explain the observed one,
while observationally, the 21$\mum$ feature in all sources
has an (almost) identical intrinsic spectral profile
(see Volk et al.\ 1999).

With $\Eobs \approx 1.0\times 10^{36}\erg\s^{-1}$ 
for the 21$\mum$ emission feature of HD\,56126 
(Hony et al.\ 2003)
and $M_{\rm H}\approx$\,0.20\,${\rm M_\odot}$
for the circumstellar envelope of HD\,56126
(Zhang et al.\ 2009b),\footnote{%
  The mass of the circumstellar envelope of HD\,56126
  is not precisely known.
  Hony et al.\ (2003) derived 
  a circumstellar envelope mass of
  $M_{\rm H}$\,$\sim$\,0.16--0.44\,${\rm M_\odot}$,
  depending on the assumed gas-to-dust ratio ($\simali$220--600).
  Meixner et al.\ (2004) derived
  a much smaller mass 
  of $M_{\rm H}\sim$\,0.059\,${\rm M_\odot}$
  based on the CO J\,=\,1--0 line emission images.
  Zhang et al.\ (2009b) derived 
  $M_{\rm H}\sim$\,0.20\,${\rm M_\odot}$,
  a value which is intermediate between the estimation of
  $M_{\rm H}$\,$\sim$\,0.16--0.44\,${\rm M_\odot}$ 
  of Hony et al.\ (2003). 
  We therefore adopt $M_{\rm H}\sim$\,0.20\,${\rm M_\odot}$.
  }
we calculate a total FeO mass of
$M({\rm FeO})\approx 2.38\times10^{29}\g$
which is required to account for
the 21$\mum$ emission feature.
The Fe abundance required to be locked up in FeO
is $\left[{\rm Fe/H}\right]_{\rm FeO}\approx 8.31\times10^{-6}$,
exceeding what is available in HD\,56126
($\left[{\rm Fe/H}\right]_\star\approx 3.24\times10^{-6}$;
van Winckel \& Reyniers 2000) 
by a factor of $\simali$2.6.\footnote{%
  Zhang et al.\ (2009a) derived
  $\left[{\rm Fe/H}\right]_{\rm FeO}\approx 5.76\times10^{-7}$
  by assuming FeO emits at the peak temperature 
  which it reaches upon absorption of a typical stellar photon.
  They adopted a Debye temperature of 
  $\DT = 650\K$ which is higher than 
  $\DT \approx 430\K$ derived here by 
  a factor of $\simali$1.5.
  For a given stellar photon, 
  this would overestimate the FeO temperature
  by a factor of $\simali$1.5$^{3/4}$\,$\approx1.36$
  and therefore underestimate the required
  FeO mass by a factor of $\simali$1.36$^6$\,$\approx6.2$.
  }

\begin{figure}[ht]
\begin{center}
\epsfig{
        file=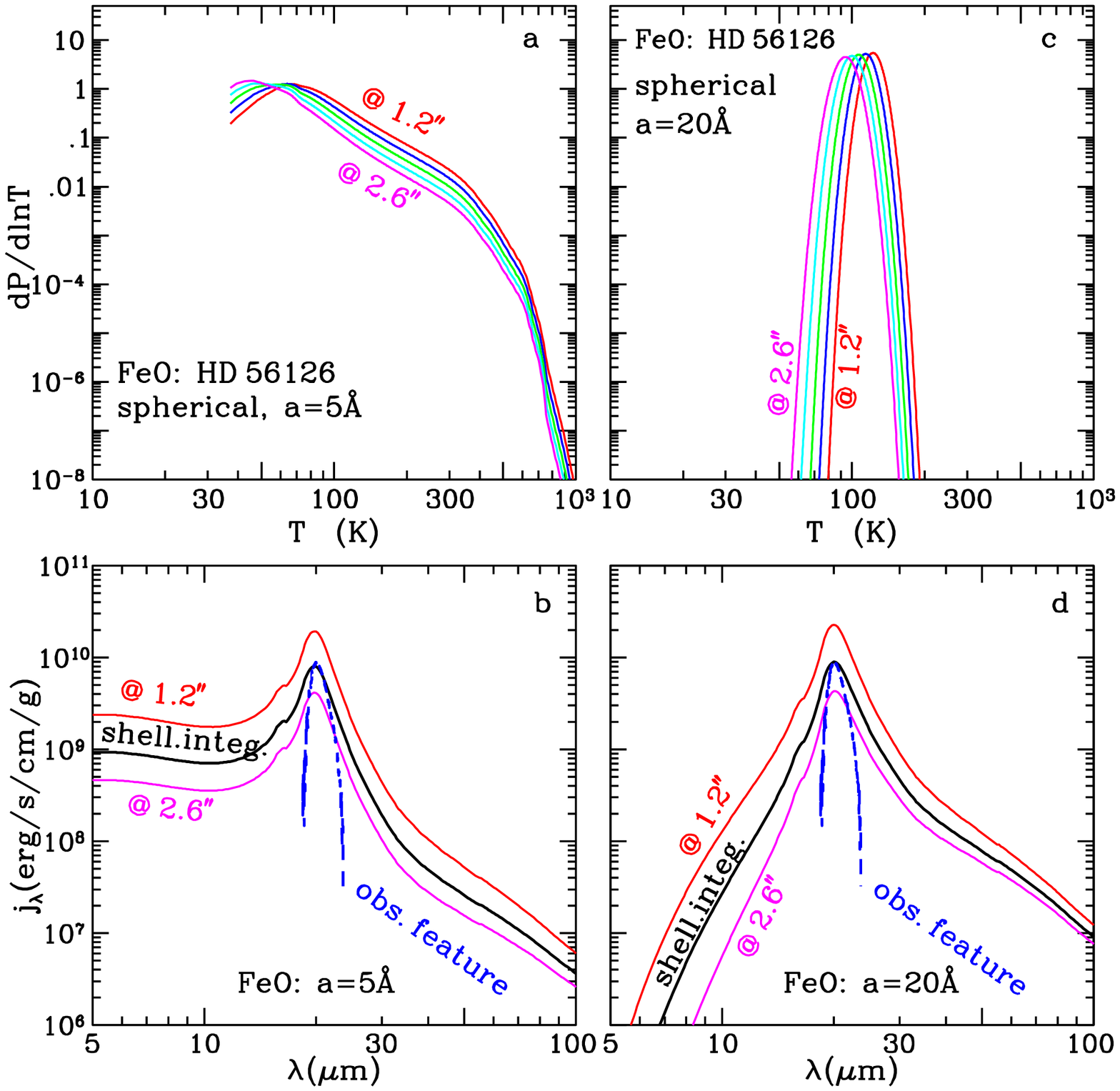,
        width=\figwidthL,angle=0}
\end{center}\vspace*{-1em}
\caption{\footnotesize
        \label{fig:irem5A20A}
        Same as Figure~\ref{fig:irem10A}
        but for FeO of $a$\,=\,5$\Angstrom$ (left panels: a,\,b)
        and $a$\,=\,2\,nm (right panels: c,\,d).
        }
\end{figure}

Similarly, we have calculated the $P(r,a,T)$ distribution
functions and the IR emission spectra for FeO nanoparticles
of $a$\,=\,5$\Angstrom$ 
and $a$\,=\,2\,nm (see Figure~\ref{fig:irem5A20A}).
Compared to that of FeO of $a$\,=\,1\,nm,
the $P(r,a,T)$ distribution function for $a$\,=\,5$\Angstrom$ 
is much broader and is appreciably large even at $T>300\K$.
This is because for the $a$\,=\,5$\Angstrom$ FeO
(with only $\simali$50 atoms), its heat capacity is
only 1/8 of that of the $a$\,=\,1\,nm FeO and therefore
it can be heated to a much higher temperature by the same 
stellar photon. On the other hand, with a cross section 
only 1/4 of that of the $a$\,=\,1\,nm FeO, its photon absorption
rate is also just 1/4 of that of the $a$\,=\,1\,nm FeO,
and therefore, the single-photon heating effect becomes
more pronounced for the $a$\,=\,5$\Angstrom$ FeO.
In contrast, the $a$\,=\,2\,nm FeO, with a heat capacity
eight times that of the $a$\,=\,1\,nm FeO, 
the $P(r,a,T)$ distribution function is more narrowly peaked
at its ``equilibrium'' temperature of $T\approx 120\K$
(see Figure~\ref{fig:irem5A20A}), implying that the single-photon
heating effect becomes less significant for larger FeO dust.
The criterion for single-photon heating will be discussed
in detail in \S\ref{sec:discussion}.

The model 21$\mum$ feature from the $a$\,=\,5$\Angstrom$ FeO,
with a FWHM $\gamma_{\rm 21\mu m}\approx 3.6\mum$, is also too
broad compared to that of the observed feature.
To account for the 21$\mum$ emission feature observed 
in HD\,56126, the $a$\,=\,5$\Angstrom$ FeO requires
$M({\rm FeO})\approx 2.78\times10^{29}\g$ and
$\left[{\rm Fe/H}\right]_{\rm FeO}\approx 9.70\times10^{-6}$.
For the $a$\,=\,2\,nm FeO, the corresponding numbers
are $\gamma_{\rm 21\mu m}\approx 3.7\mum$, 
$M({\rm FeO})\approx 2.54\times10^{29}\g$, and
$\left[{\rm Fe/H}\right]_{\rm FeO}\approx 8.87\times10^{-6}$.

\begin{figure}[ht]
\begin{center}
\epsfig{
        file=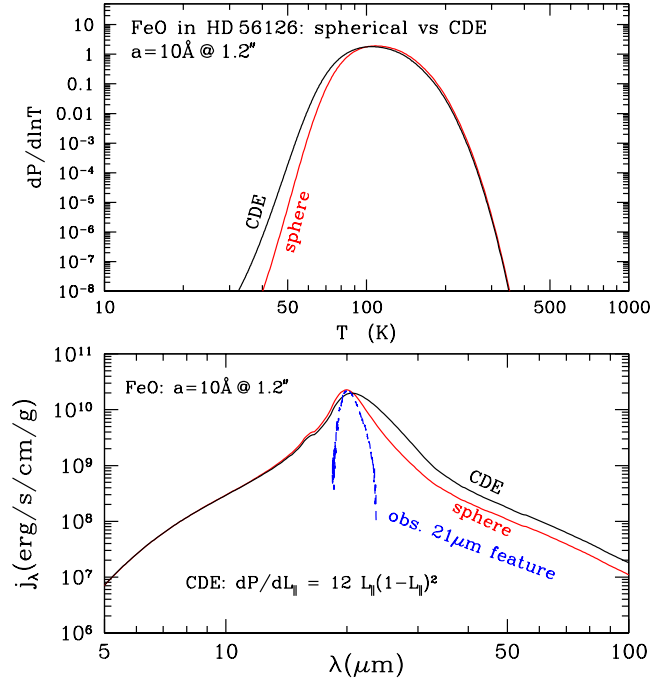,
        width=\figwidth,angle=0}
\end{center}\vspace*{-1em}
\caption{\footnotesize
        \label{fig:irem10Acde}
        Upper panel: the temperature probability
        distribution functions of 
        (1) spherical FeO (red) and
        (2) spheroidal FeO with a distribution 
        of shapes (black) 
        of radius $a$\,=\,1\,nm in HD\,56126 
        at 1.20$^{\prime\prime}$
        from the star.
        Lower panel: the IR emissivity per unit mass
        of (1) spherical FeO (red) and 
        (2) spheroidal FeO with a distribution 
        of shapes (black) of radius $a$\,=\,1\,nm at 
         1.20$^{\prime\prime}$ from the star.
         Also shown is the observed 21$\mum$ emission feature 
         of HD\,56126 (blue dashed; Volk et al.\ 1999) 
         which is scaled to that of spherical FeO.
         The 21$\mum$ feature arising from 
         the spherical $a$\,=\,1\,nm FeO
         (with a FWHM of $\gamma_{\rm 21\mu m}$\,$\approx$\,3.5$\mum$)
         is much narrower than that of spheroidal FeO
         with a distribution of shapes
         (for which the FWHM is 
          $\gamma_{\rm 21\mu m}$\,$\approx$\,5.6$\mum$).
         }
\end{figure}

\section{Discussion\label{sec:discussion}}
We have seen in \S\ref{sec:results} that FeO nanoparticles
emit strongly at 21$\mum$. However, the model 21$\mum$ feature
(with a FWHM of $\gamma_{\rm 21\mu m}$\,$\approx$\,3.6--3.7$\mum$)
is much broader than the observed feature
(with a FWHM of $\simali$2.2$\mum$).
The Fe abundance required to be locked up in FeO 
($\left[{\rm Fe/H}\right]_{\rm FeO}$)
exceeds the available abundance 
$\left[{\rm Fe/H}\right]_\star$
by a factor of $\simali$2.6--3.
These results are obtained for spherical grains.
For nonspherical grains, the model 21$\mum$ feature
would be even broader. 
We have calculated the IR emission spectra for 
FeO nanoparticles with a distribution of spheroidal shapes with 
$dP/dL_{\parallel} = 12 L_{\parallel}[1-L_{\parallel}]^2$ 
(Ossenkopf et al.\ 1992) 
where $0<L_{\parallel}<1$ is the so-called ``depolarization factor'' 
parallel to the grain symmetry axis (for spheres $L_{\parallel}$=1/3); 
this shape distribution peaks at spheres and drops to zero for 
the extreme cases $L_{\parallel}\rightarrow 0$ (infinitely thin needles)
or $L_{\parallel}\rightarrow 1$ (infinitely flattened pancake). 
As shown in Figure \ref{fig:irem10Acde}, 
the model 21$\mum$ emission feature arising from
FeO nanoparticles with such a distribution of
spheroidal shapes is much broader than that of spherical FeO.

In calculating the absorption cross sections of FeO,
we use the optical constants of $T$\,=\,100$\K$.
As shown in Figure~9 of Posch et al.\ (2004),
the absorption profile of the 21$\mum$ band 
broadens if one adopts the optical constants 
of $T$\,=\,200, 300$\K$,
while essentially it remains unchanged 
from  $T$\,=\,100$\K$ to $T$\,=\,10$\K$.
Although small, $P(r,a,T)$ is positive at $T>200\K$
for $a\simlt1\nm$ 
(see Figures~\ref{fig:irem10A},\ref{fig:irem5A20A}).
Therefore, if we adopt the ($\indxre$, $\indxim$) data
of $T\simgt200\K$ for FeO warmer than $200\K$, 
we would expect a broader 21$\mum$ emission feature.
Nevertheless, we note that the 21$\mum$ feature of 
nano FeO calculated in this work is based on 
the dielectric functions of bulk FeO material
(see \S\ref{sec:opct}). It is not clear how the dielectric
functions near the 21$\mum$ resonance wavelength range
will be affected when FeO becomes nano-sized.

For a small {\it metallic} grain, the imaginary part
of its dielectric function is expected to be larger 
compared to that of its bulk counterpart, as a consequence 
of the so-called {\it electron mean free path limitation} effect
(see \S6 in Li 2004). 
For FeO which is a {\it semiconductor} 
(Seagle et al.\ 2009, Schrettle et al.\ 2012),
the number density of free charge carriers is lower
than that of metals by several orders of magnitude
(see Henning \& Mutschke 1997). Therefore, the small
size effect on the dielectric function caused by 
the limitation of the electron mean free path     
is expected to be less significant for FeO than for metals.

We also note that the thermal properties of nano FeO may 
differ from that of bulk material (see \S\ref{sec:heatcap}).
The specific heats of some small metal particles are
reported to be strongly enhanced over their bulk values
(see \S6 in Li 2004). It is not clear how the Debye temperature
$\DT$ of nano FeO would compare to that of bulk FeO.
If nano FeO is like palladium 
(with $\DT\approx 273\K$ for bulk palladium 
and  $\DT\approx 175\K$ for nano palladium of $a$\,=\,1.5\,nm),
the stochastic heating effect 
will be less substantial.\footnote{%
    With a smaller Debye temperature $\DT$,
    the specific heat $C(T) \propto \left(T/\DT\right)^{3}$
    increases. Therefore, upon absorption of a photon of
    energy $h\nu$, the temperature rise $\Delta T$
    (from the initial temperature $T_i$ 
    to the final temperature $T_f$) will be less:
    $\Delta T = T_f - T_i$, $\int_{T_i}^{T_f} C(T) dT =h\nu$.
    }
But we also note that the specific heats 
of nano silicon crystals
and nanocrystalline diamonds do not 
differ much from their bulk values
(see \S6 in Li 2004). 

Finally, we note that ideally, we should have justified why, 
in the first place, it is necessary to consider the 
stochastic heating of nano FeO. 
It is now well recognized that
a grain undergoes stochastic heating 
by single stellar photons if
(i) its heat content is smaller than or comparable 
    to the energy of a single stellar photon 
    (Greenberg 1968), and
(ii) the photon absorption rate is smaller than
the radiative cooling rate (Draine \& Li 2001).

The photon absorption time scale --- 
the mean time $\tau_{\rm abs}$ between photon absorptions
--- for a nano FeO of size $a$ is given by
\beq
\tau_{\rm abs}^{-1} \equiv 
\int_0^\infty C_{\rm abs}(a,\lambda)
\frac{cu_\lambda}{hc/\lambda} d\lambda ~~,
\eeq
where $c$ is the speed of light, 
$h$ is the Planck constant, and 
$u_\lambda$ is the starlight energy density.
The mean photon energy 
$\langle h\nu\rangle_{\rm abs}$ 
absorbed by the FeO dust of size $a$ is
\beq
\langle h\nu \rangle_{\rm abs} \equiv
\tau_{\rm abs}\int_0^\infty C_{\rm abs}(a,\lambda) 
c u_\lambda d\lambda ~~.
\eeq
The radiative cooling time for a FeO grain
of size $a$ containing a vibrational energy 
of $\langle h\nu\rangle_{\rm abs}$ is
\beq
\tau_{\rm cooling} \approx 
\frac{\langle h\nu\rangle_{\rm abs}} 
{\int_{\lambda_{\rm min}}^\infty C_{\rm abs}(a,\lambda) 
4\pi B_\lambda(T_p) d\lambda} ~~,
\eeq
where $\lambda_{\rm min}\equiv hc/\langle h\nu\rangle_{\rm abs}$
and $T_p$ is determined by its heat content 
$E(T_p) = \int_{0}^{T_p} C(T) dT 
= \langle h\nu\rangle_{\rm abs}$. 

\begin{figure}[ht]
\begin{center}
\epsfig{
        file=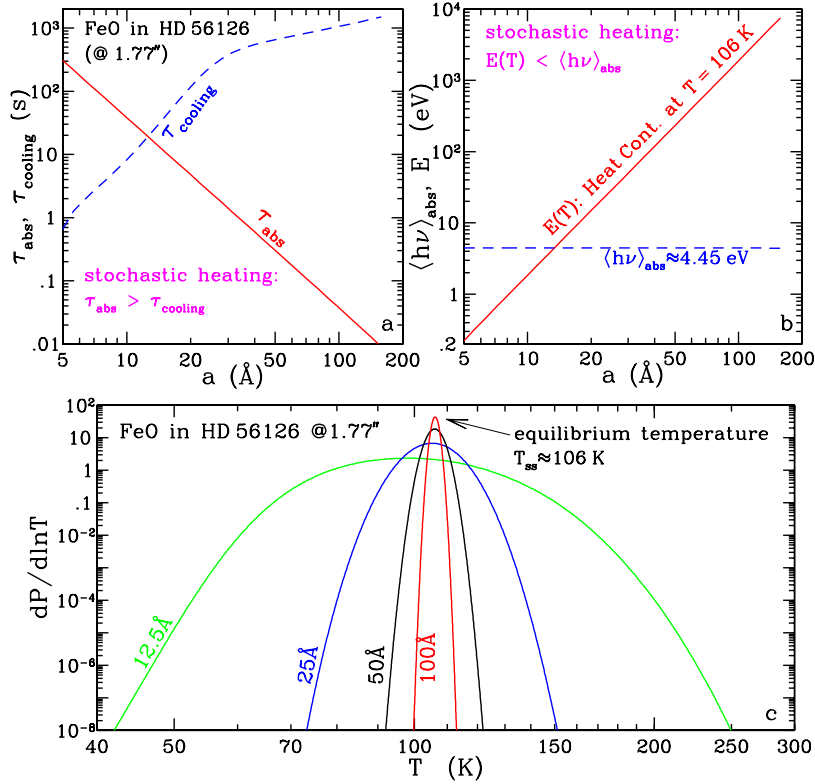,
        width=\figwidthL,angle=0}
\end{center}\vspace*{-1em}
\caption{\footnotesize
        \label{fig:tauabs}
        Upper left panel (a): 
        Comparison of 
        the radiative cooling time scale 
        $\tau_{\rm cooling}$ for FeO dust
        containing a vibrational energy of
        $\langle h\nu\rangle_{\rm abs}$
        with the photon absorption time scale 
        $\tau_{\rm abs}$ for FeO dust at
        $r=1.77^{\prime\prime}$ in HD\,56126.  
        Upper right panel (b): 
        Comparison of the mean absorbed photon energy 
        $\langle h\nu\rangle_{\rm abs}\approx4.45\eV$
        with the heat content 
        $E(T_{\rm ss}) = \int_{0}^{T_{\rm ss}} C(T) dT$ 
        at $T_{\rm ss}=106\K$, the ``equilibrium'' temperature
        which would be attained by FeO dust
        of $a\simgt5\nm$ at $r=1.77^{\prime\prime}$
        in HD\,56126.
        Bottom panel (c): The temperature probability
        distribution functions of FeO dust 
        at $r=1.77^{\prime\prime}$ in HD\,56126.
        }
\end{figure}

For nano FeO of $a<30\nm$ 
in the dust shell around HD\,56126, 
the mean absorbed photon energy is
$\langle h\nu\rangle_{\rm abs}\approx 4.45\eV$,
independent of dust size $a$.
This is because in the UV/visible wavelength range
nano FeO is in the Rayleigh regime
and its absorption properties 
are independent of size $a$.
In Figure~\ref{fig:tauabs}a we show $\tau_{\rm abs}$
and $\tau_{\rm cooling}$ of FeO dust
in the HD\,56126 dust shell,
at a distance of $r=1.77^{\prime\prime}$ from 
the central star
which is intermediate between the inner boundary
and the outer boundary of the shell.
It is apparent that $\tau_{\rm abs}\propto a^{-2}$ 
rapidly decreases with $a$, while $\tau_{\rm cooling}$
increases with $a$.
It is clear that FeO dust of $a<1.2\nm$,
with $\tau_{\rm abs} > \tau_{\rm cooling}$,
will be subject to substantial 
temporal fluctuations in temperature.
For FeO dust of $a\simgt5\nm$,
with $\tau_{\rm abs} \ll \tau_{\rm cooling}$,
will attain an equilibrium temperature.

In Figure~\ref{fig:tauabs}b we compare 
the mean absorbed photon energy 
$\langle h\nu\rangle_{\rm abs}$
with the heat content 
$E(T_{\rm ss}) = \int_{0}^{T_{\rm ss}} C(T) dT$ 
at $T_{\rm ss}=106\K$, the ``equilibrium'' 
or ``steady-state'' temperature
which would be attained by FeO dust
of $a\simgt5\nm$ at $r=1.77^{\prime\prime}$.
Again, it is seen that for FeO dust of 
a couple of nanometers in size,
its heat content is comparable or smaller
than the energy of a single stellar photon
and its temperature will be appreciably raised
upon absorption of an individual photon,
while for FeO dust of $a\simgt5\nm$,
its heat content is much larger than
the photon energy and therefore the absorption
of a single photon will not change its temperature.
This also justifies the necessity to consider
the stochastic heating of nano FeO of $a<5\nm$.

In Figure~\ref{fig:tauabs}c we show
the temperature probability distribution functions
$dP/d\ln T$ for nano FeO of $a$\,=\,1.25, 2.5, 5, 10$\nm$
at $r=1.77^{\prime\prime}$.
These results confirm the conclusions drawn from
the general considerations discussed in 
Figure~\ref{fig:tauabs}a,b:
for FeO dust of $a\simgt5\nm$,
with $\tau_{\rm abs}\ll \tau_{\rm cooling}$
and $E(T_{\rm ss}) \gg \langle h\nu\rangle_{\rm abs}$
at $T_{\rm ss}\approx 106\K$, 
the temperature distribution function
$dP/d\ln T$ is like a delta function,
implying that it will attain an equilibrium 
temperature of $T_{\rm ss}\approx 106\K$.
According to Posch et al.\ (2004),
the size of the FeO dust which could form and survive
in the C-rich shells around the 21$\mum$ sources
should not exceed $\simali$1\,nm. 
FeO dust this small will have 
$\tau_{\rm abs} > \tau_{\rm cooling}$  
(see Figure~\ref{fig:tauabs}a),
and $E(T) < \langle h\nu\rangle_{\rm abs}$
at $T_{\rm ss}\approx 106\K$
(see Figure~\ref{fig:tauabs}b),
and therefore will be stochastically heated
by single stellar photons.

To summarize, we have examined the hypothesis of
FeO nanoparticles as a carrier of the mysterious 
21$\mum$ emission feature seen in C-rich PPNe.
The temperature probability distribution functions
and the resulting IR emission spectra have been calculated
for these stochastically-heated nano-sized grains.
We find that they emit too broad a 21$\mum$ feature 
to explain the observed one and the Fe abundance required
to be locked up in FeO exceeds what is available.
This, combined with the special conditions required
for the formation of FeO nanoparticles in C-rich 
environments (see \S3.3 of Posch et al.\ 2004, Duley 1980),
leads us to conclude that probably FeO nanoparticles 
are not responsible for the 21$\mum$ feature.

\acknowledgments
We thank J. Gao and K. Zhang for their kind help
in preparing for this manuscript.
We thank K. Volk who kindly provided us the 21$\mum$
emission spectra of the four PPNe shown in Figure~\ref{fig:Obs}.
We thank S. Hony, A.P. Jones, C. Koike, P. Miceli, 
and the anonymous referee for helpful suggestions.
We are supported in part by
NSF AST-1109039, NNX13AE63G, 
NSFC\,11173007, NSFC\,11173019, NSFC\,11273022,
and the University of Missouri Research Board.

\end{document}